# STUDY AND PERFORMANCE EVALUATION OF SECURITY-THROUGHPUT TRADEOFF WITH LINK ADAPTIVE ENCRYPTION SCHEME

Poonam Jindal[1] and Brahmjit Singh[2]

Department of Electronics and Communication Engineering, NIT Kurukshetra
*poonamjindal81@yahoo.co.in*

## ABSTRACT

*With the ever increasing volume of information over wireless medium, security has assumed an important dimension. The security of transmitted data over a wireless channel aims at protecting the data from unauthorized intrusion. Wireless network security is achieved using cryptographic primitives. Some properties that give encryption mechanism their cryptographic strength also make them very sensitive to channel error as well. Therefore, security for data transmission over wireless channel results in throughput loss. Trade-off between security and throughput is always a major concern in wireless networks. In this paper, a Link Adaptive Encryption scheme is evaluated that adapts to channel variations and enhances the security level of WLANs without making any compromise with the network performance. Numerical results obtained through simulation are compared with the fixed block length encryption technique in two different modes of operation- Electronic Code Book (ECB) & Cipher Block Chaining (CBC). Optimal block length is also computed, which is assumed to be the effective strength of the cipher. It has been observed that security attained with link adaptive scheme operating in ECB mode of cipher is a better solution for security and throughput trade-off. However, it is found that if computational security is a major concern, link adaptive scheme in CBC mode should be preferred.*

## 1. INTRODUCTION

There is a tremendous growth of wireless communication services over the last decade. The security concerns are becoming more serious with the growth of wireless networks. As more people access critical information, and consumers begin to do their business and banking on mobile wireless devices wireless security has moved to the forefront. But there are certain attacks that exist on wireless networks and these can be classified as active and passive attacks [1]. Active attacks are those in which the transmitted data can be modified by an intruder and in passive attacks transmitted data can only be read [2]. Cryptographic primitives such as encryption and authentication are used to secure data over the wireless channel. To scramble a message in such a way that intervener could not see/modify the data is known as encryption and to identify the person who is communicating with you is authentication.

To prevent the information from adversary attacks encryption algorithms have been designed. The cipher is made to satisfy several properties including avalanche effect. Avalanche effect can be defined as a single bit change in a plaintext or the key must result in significant changes to the ciphertext. So there should not be any resemblance between the transmitted plaintext and received ciphertext [2]. But this property of block ciphers leads to severe error propagation i.e. the single bit error in the received encrypted block will lead to propagate an error in every bit of decrypted block with half of the probability, which results in a loss of throughput. This leads to a tradeoff between security and throughput towards making the wireless networks secure. In this paper, we analyze security-throughput trade-off in WLANs.

It has been observed that the quality of wireless channel varies due to time varying path loss and multipath fading phenomenon. The author in [3] proposes that in order to achieve the capacity of wireless communication systems, channel adaptive resource assignment techniques are required. To achieve this channel adaptive transmission rate and power assignments are proposed in [4]. Because the security algorithms are sensitive to the errors induced by the channel, the transmission overheads incurred by these algorithms also vary with time [2].

WEP was the first security protocol [5], designed to cover security requirements in wireless environment, which provides authentication, data integrity and privacy. WPA contributes to the increase of wireless communication protection by Wi-Fi standard through increased level of data protection, access control and integrity [6]. Impact of different implementations of encryption techniques used by two security protocols, namely Wired Equivalent Privacy (WEP) and Wi-Fi Protected Access (WPA) on the throughput over WLAN IEEE 802.11g have been discussed in [7]. The results show that within the same access point range the security adds moderate degradation in the throughput that may affect some applications over both infrastructure and ad hoc WLANs. The security strength of standard security protocols used in WLANs and their energy consumption is analyzed in [8]. The impact of security on latency in WLAN 802.11b is studied in [9]. The average authentication delay increases with the security level of the configuration. The effect of multiple security mechanisms on the performance of multi-client congested and uncongested networks is presented in [10].

It is established that there is always a tradeoff between security and the performance of wireless networks. As we increase the security there is degradation in the system performance. The problem of this security–throughput tradeoff is a major research issue in the implementation of security algorithms in wireless networks.

In traditionally designed security/encryption algorithms, the inclusion of bit errors during the transmission of data/information over the channel was not taken into consideration and this has been considered an orthogonal problem which is required to be handled by efficient coding and modulation techniques. The present and future wireless communication systems can be benefited from an encryption designs that considered the channel quality and make it possible to achieve a desirable tradeoff between the security and performance of wireless network [11].

The tradeoff between security and throughput for given channel conditions has been shown in fig.1 with the assumption that block length is always equal to or less than the key length [12].

In a WLAN, normalized throughput may be defined as

$$(1 - P_i)^N \qquad (1)$$

Where: $P_i$ bit error probability and N is the encryption block length. Normalized security may also be computed as $Log_2 N$. Fig.1 shows the security and throughput as a function of block length. The set of block lengths used in the plot includes a single bit from 16 bits to 256 bits in increments of 16 bits. If N is a block length, then number of computations required by an eavesdropper will be $2^N$. The security of a cipher can be defined as $2^N$. It is obvious that both security and throughput are inversely proportional to each other.

The problem of tradeoff between security and throughput has been optimized using link adaptive technique [12]. In this technique, we utilize the mean channel opportunity which implies the time duration when signal to noise ratio (SNR) is high or corresponding bit error rate (BER) is low.

In link adaptive encryption scheme, when SNR is high, data is encrypted in larger blocks. In case, SNR is low, data is encrypted with smaller block size. With high SNR, larger block length increases the security level of the system and due to less probability of error the overall throughput also increases. In case of low SNR or high BER, it is preferred to encrypt the data with smaller block length, leading to enhanced throughput. This in turn results in improved security-throughput tradeoff. In case of low SNR, the ability of cryptanalyst to decrypt the cipher is also less and thus maintains the security of the cipher. In the process, feature of variable encryption block length has been exploited. In the security-throughput tradeoff analysis with link adaptive technique, it is assumed that the channel states are known for the transmission period of the message. Further, Advanced Encryption Standard (AES) (block cipher) operating in both ECB and CBC modes of cipher has been adopted as an encryption algorithm in the present work.

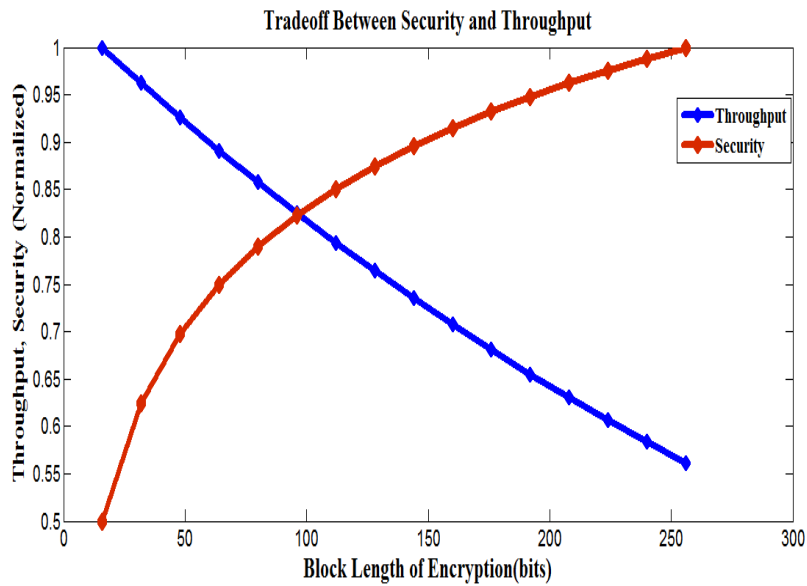

Figure.1 Security-Throughput (Normalized) Tradeoff as a Function of Block Length at

Channel Bit Error Probability, $P_i$ =.0024

To quantify the amount of security provided by a cipher is a complex problem. One method to measure the level of security provided by a cipher against cryptanalysis is the computation operation required by the adversary to crack the ciphertext. The level of security can be quantified relative to the strength of the intruder present in the environment. Intruder's strength can be modeled using probability distribution function under the assumption that the ability of an adversary to crack the ciphertext becomes less probable as the computational complexity of cipher increases. The probabilistic adversary strength model has been introduced in [12].

The probabilistic model calculates the maximum possible block length the adversary can crack. This probabilistic model has been used in the present work to evaluate the cipher strength. Cipher strength is determined in the terms of the adversary's capability to crack a cipher with a block length of N bits. This fact implies that adversary is capable to crack any block length less than or equal to N bits.

In view of the literature survey discussed above, it is established that the problem of security-throughput tradeoff with link adaptive encryption scheme has been well studied with AES operating in Electronic Code book (ECB) mode [12]-[14]. However, ECB mode is known to be a less secure mode of cipher and CBC mode is computationally more secure as compared to ECB mode of cipher. In this paper, we present and compare the results on the performance of link adaptive encryption scheme in both ECB and CBC mode. We have utilized optimal strength of the cipher in our system modeling which to the best of our knowledge was not taken into consideration in the literature available.

The rest of the paper is organized as follows. Brief introduction to the different modes of cipher and Block cipher (AES) is given in section 2. Security measurement of the cipher is described in section 3. Simulation of system model with fixed length encryption and channel/link adaptive scheme is discussed in section 4. Simulation results for channel adaptive scheme are discussed in section 5. Conclusion and future work is outlined in section 6.

## 2. SECURITY ALGORITHMS

The data transmitted over a wireless network is highly insecure due to the open nature of wireless medium. To protect the data from eavesdropping some security measures are required. One method to provide security is the use of encryption algorithms in which the plaintext or the input data is converted into a ciphertext which is a scrambled form of input data or unintelligible form of a message. A number of schemes have been proposed for encryption [2].

Ciphers are classified into symmetrical cipher and asymmetrical cipher. Symmetric key ciphers are classified into block cipher and stream cipher. The cipher that we have used in our work is a symmetrical block cipher in which data is processed in blocks i.e. one block of elements at a time is taken as input, producing an output block for each input block with the same random key shared by both the sender and receiver. Ciphers such as DES (Data Encryption Standard), 3DES, AES (Advanced encryption Standard), IDEA (International Data Encryption Algorithm) are different available block ciphers [15]. Till date AES is known to be the most secure encryption algorithm with the feature of variable block length [16].

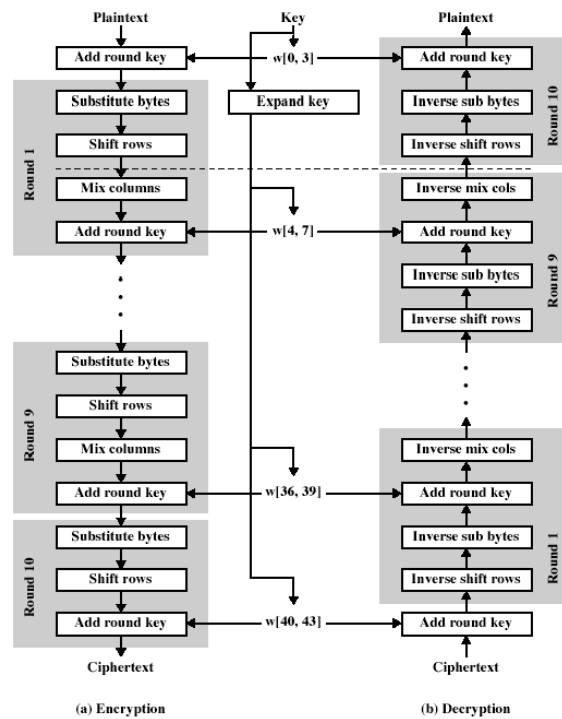

Figure 2: AES Encryption [1]

## 2.1 AES

The Advanced Encryption Standard (AES) standard became effective on May 26, 2002 by NIST to replace DES. The AES is a symmetric block cipher that encrypts and decrypts 128-bit blocks of data. Rijndael consists of a number of rounds; each round performs number of transformations on a state, and makes use of a round key which is derived from the encryption key. The number of rounds depends on the block and key sizes.

An encryption of a block starts with a transformation 'AddRoundKey', which is followed by a number of regular rounds, and ends with a special final round. Special final round is different from the regular rounds. This difference in the final round has nothing to do with security, but makes the encryption reversible and makes the decryption possible. In AES each transformation is performed in a state which can be pictured as a rectangular array of bytes. It consists of four rows and a number of columns defined by the block size in bytes divided by four. A block size of 128 bits would require a state of four rows.

AES encryption and decryption are shown in fig 2. There are four transformations in one round:

1. AddRoundKey
2. SubBytes
3. ShiftRows
4. MixColumns

A key size of 128 would require 10 numbers of rounds. All the nine rounds would perform all the four transformations but the final round would have only three steps. MixColumn transformation is not present in the final round.

The Roundkeys are made by expanding the encryption key into an array holding the RoundKeys one after another. The expansion works on words of four bytes. The Rijndael proposal for AES defined a cipher which can operate on different block lengths and key lengths such as 128, 160, 192, 224, and 256 bits. This property of AES is exploited in the channel adaptive scheme.

## 2.2 Modes of Operation

Block ciphers are basic building blocks for providing data security. To enhance the security of ciphers further, block ciphers are used in different modes of operation. These are:

1. Electronic Code Book (ECB)
2. Cipher Block Chaining (CBC)
3. Cipher Feedback (CFB)
4. Output Feedback (OFB)
5. Counter (CTR)
6. Galois/Counter Mode (GCM)

Some other modes of operations have also been discussed in NIST [17]-[19]. Authentication modes include version of CBC-MAC (XCBC), Randomized CBC-MAC (RMAC) and Parallelizable Message Authentication Code (PMAC), and authenticated encryption modes: OCB, XCBC-XOR, XECB-XOR; IAPM and IACBC.

In the present work trade-off analysis has been done with ECB and CBC modes of cipher. In ECB there is a parallel processing of data i.e. every plaintext block is independently encrypted to a cipher text block. So there is no error propagation in ECB mode. In CBC mode, a given plaintext block is XORed with the ciphertext block of the previous stage before encryption. So if there is an error in one ciphertext block, error will be propagated to multiple decrypted plaintext blocks.

## 3. MEASUREMENT OF THE SECURITY OF A CIPHER

One way to measure the security of the cipher is to measure the amount of work required by the adversary to crack a ciphertext. It is required to make the system computationally secure. A cipher is said to be computationally secure if the cost of breaking the information is more than the value of the information or if the time required in breaking the ciphertext is more than useful lifetime of the information being sent. It has been assumed in the present work that ability of the adversary to break the cipher becomes less probable if there is an increase in block length, key length, diffusion etc. To measure the strength of the cipher we have to measure the strength of the adversary which is also a vulnerability of the message. The strength of the adversary can be defined as the maximum block length the adversary can crack or vulnerability of a message can be defined as a fraction of the total message being successfully decrypted by the intruder/adversary. We have calculated the resistance of cipher against brute force attack by calculating the optimal block length. Optimal block length is the length of plaintext/ ciphertext in bits to which adversary can easily break. For example if the optimal block length that we obtain is 160 bits, it means that the intruder can crack the encryption blocks which use 160 bits or lesser.

The vulnerability $\Phi$ of the message that is the fraction of the total message being successfully decrypted by the intruder/adversary is given by (2) [13].

$$\phi = \sum_{k=1}^{K} x_k P(x_k) \tag{2}$$

Where, $x_k$ is the leakage of the fraction of message and $P(x_k)$ is the probability of exposing a fraction $x_k$ of the total message.

## 4. SYSTEM MODEL

Channel adaptive scheme is the scheme which adapts the channel variations. This scheme is the way to optimize the tradeoff between security offered and throughput loss due to the cipher. In the simulation, the encrypted data is transmitted over a wireless channel. Then the transmitted data is received at the receiver end, and then decrypted back to achieve a desired plaintext.

### 4.1 Optimization of Security Throughput Tradeoff

AES supports different block lengths. It is assumed that no significant overheads occur while changing the block length of the cipher. Further, key length is always equal to the block length. Optimization is achieved under the assumption of exact channel knowledge of the channel state in terms of SNR and BER. Avalanche effect causes errors within an encryption block. A single bit error in the received encrypted block will cause the loss of entire block due to the error propagation after decryption. It is, therefore, desirable to increase the overall throughput of the system while not making any compromise with the required security level [13]. The throughput per block and, in turn for a frame, for $P_i \ll 1$ and for given fixed $N_i$ may be defined as:

$$R_i(1 - N_i P_i) \tag{3}$$

Where, $R_i$ and $P_i$ are the selected transmission rate and the channel bit error probability. The throughput of the message can be expressed as:

$$T = \frac{1}{nR_{max}} \sum_{i=1}^{n} R_i(1 - N_i P_i) \tag{4}$$

The throughput is normalized by the maximum transmission rate $R_{max}= \max\{R_i\}$. In the present work security quantification is performed with brute force attack model. To choose the encryption block length based on the channel condition and to obtain the desired security level is the key to this optimization process. The strategy for the choice of optimal block length depends on the channel conditions. Further there is a requirement for the receiver to know the encryption block length that has been used by the sender during the transmission of each frame. The simple approach that can be used to achieve this is to include the block length information as a clear payload in the frame. Another approach that can be used is that the receiver can compute it from the security constraints and the channel states during the reception of the frame. This can be achieved as the security constraints are agreed upon a priori, and receivers, in general, have the capability to estimate the forward channel. The link adaptive scheme presented here for the optimization of the tradeoff between security and throughput highly depends on the ability to know the channel quality in terms of SNR or BER.

## 4.2 Brute Force Attack Model

Optimal block length as mentioned in section 3 also defines the effective strength of the cipher and may be given as

$$N^*_i = \frac{(\prod_{i=1}^{n} R_i P_i)^{\frac{1}{n}}}{R_i P_i} e^{(S_{req} S_{max})} \log_e(2) \tag{5}$$

Where $N^*_i$ is the optimal block length, $R_i$ and $P_i$ are respectively the transmission rate selected for the frame and channel bit error probability. $S_{req}$ is the required level of security or it is a measure of the mean of security levels achieved by the individual frames and is given by,

$$S_{req} = \frac{1}{nS_{max}} \sum_{i=1}^{n} \log_2 N_i \tag{6}$$

Where $S_{max} = \log_2 N_{max}$ ; $N_{max}$ is the maximum block length and $N_i$ is the discrete encryption block length. It is required to maximize the throughput, with subject to an overall security requirement over a finite horizon. The optimization problem as a constrained problem can be represented as [13][14]:

$$\max\{N_i\} \frac{1}{nR_{max}} \sum_{i=1}^{n} R_i(1 - N_i P_i) \tag{7}$$

As mentioned earlier, $R_i$ and $P_i$ are the selected transmission rate and the channel bit error probability. $P_i$ is a function of channel SNR and the transmission rate used for the frame. $S_{req}$ is the required level of security. The optimal block length or the optimal strength of the cipher for the fixed transmission rate (5) reduces to

$$N^*_i = \frac{(\prod_{i=1}^{n} P_i)^{\frac{1}{n}}}{P_i} e^{(S_{req} S_{max})\log_e 2} \tag{8}$$

From (8) it is found that optimal block length is inversely proportional to the channel bit error probability. This shows that in link adaptive technique, by allocating larger block lengths for better channels or vice versa, is the good approach in the case of fixed rate.

## 5. RESULTS AND DISCUSSION

The frame lengths are in general much larger than the encryption block lengths and may consist of multiple encrypted blocks. Let the message be sent by forming n frames with encrypted block length of length $N_i$ bits for the frame i=1, 2…n, where $N_i$ is selected by the channel adaptive scheme depending upon the channel conditions. Simulation results for channel adaptive scheme in ECB and CBC modes of cipher have been obtained with the help of MATLAB 7.04. The comparison of the throughput and strength of cipher (in terms of optimal block length) were observed using channel adaptive scheme and fixed block length encryption scheme (block length is fixed for whole length of the channel i.e. 128 bits) in two different modes.

### 5.1 For ECB Mode Of Cipher

Numerical results are obtained for link adaptive scheme in ECB mode of cipher. Simulation results were obtained for both channel adaptive scheme and fixed block length encryption scheme. Numerical results and graphical representation for the cipher strength with fixed and variable block length encryption are shown in table 1 and figure 3 respectively. From the results, it is observed that the maximum optimal block length with adaptive block length encryption scheme at 256 bits is 198 bits. It means that if there is any adversary present in the environment he/she can break the cipher only up to 198 bits. He/ She cannot crack the whole ciphertext. Whereas the adversary can crack the complete ciphertext with fixed block length encryption because obtained optimal block length for fixed block length encryption at 128 bits is 221 bits. Thus the security is enhanced in terms of brute force attack.

The comparative analysis of the throughput observed in simulations using link adaptive scheme (variable block length) and fixed block size encryption is shown in figure 4. Numerical results for the throughput are shown in table 2. For the illustration of optimization process the block lengths are adopted as per Rijndael cipher. The overall security requirement is set to $S_{req}$=.9759. For the link adaptive scheme, the encryption block lengths were selected from [128,160,192,224,256] (bits). The used block length for fixed block length encryption was 128 bits. Normalized values of throughput and security have been calculated at different SNRs. It has been observed that throughput is increased by 70% with link adaptive scheme as compared to fixed block length encryption scheme. From the throughput and security plots it is observed that there is an increase in throughput as well as security with channel adaptive scheme as compared to fixed block length encryption scheme.

Table 1. Cipher Strength for Fixed and Variable Block Length Encryption in ECB Mode

| SNR (dB) | Fixed Block Length Encryption | | Variable Block Length Encryption | |
|---|---|---|---|---|
| | Block Length (bit) | Cipher Strength (bit) | Block Length (bit) | Cipher Strength (bit) |
| 8 | 128 | 116 | 128 | 90 |
| 10 | 128 | 140 | 160 | 146 |
| 12 | 128 | 185 | 192 | 181 |
| 14 | 128 | 209 | 224 | 206 |
| 16 | 128 | 221 | 256 | 198 |

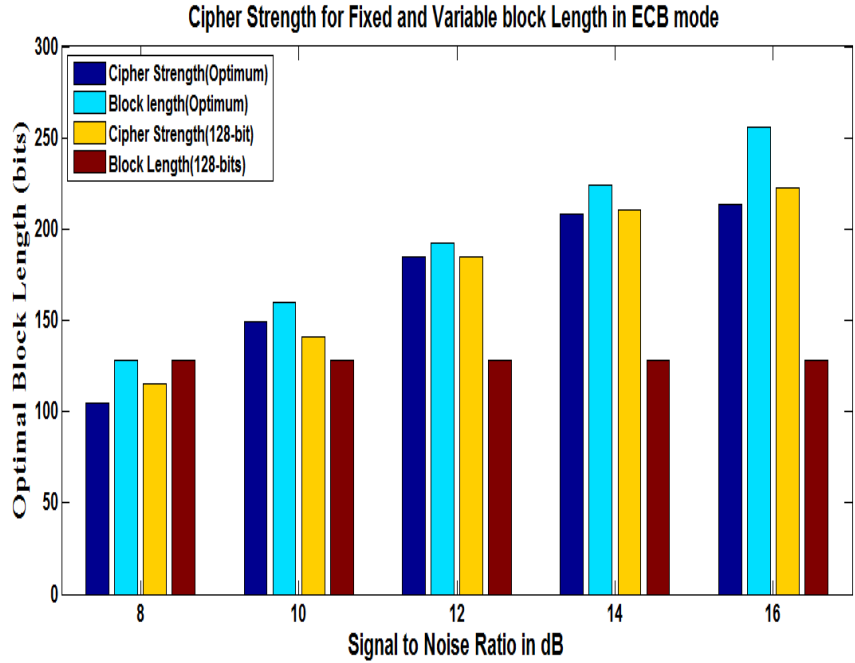

Figure 3. Cipher Strength for Fixed and Variable Block Length Encryption (Link Adaptive Scheme) in ECB Mode

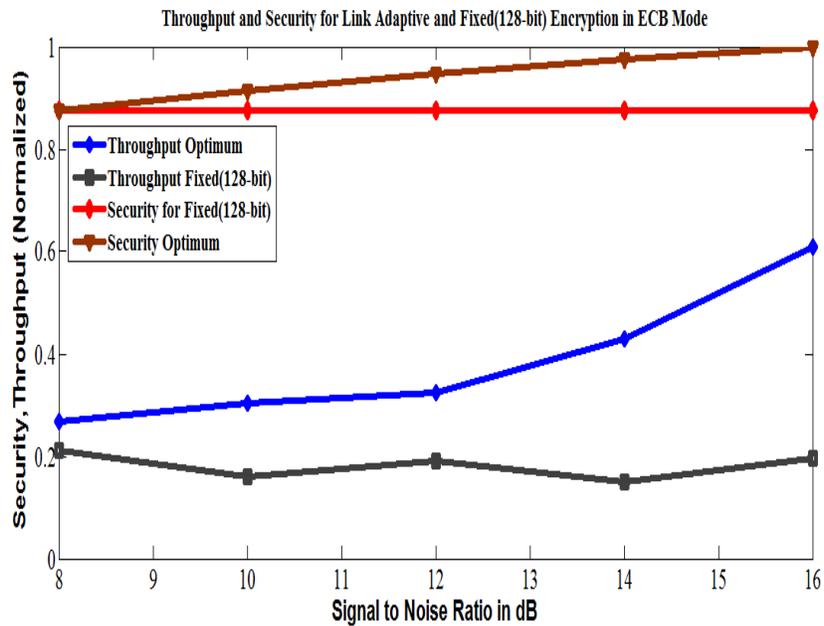

Figure 4. Normalized Throughput for Fixed and Variable Block Length Encryption $S_{req}=.97$

Table 2. Throughput for Fixed and Variable Block Length of Encryption in ECB Mode

| SNR (dB) | Fixed Block Length Encryption | | Variable Block Length Encryption | |
|---|---|---|---|---|
| | Block Length (bit) | Throughput Normalized | Block Length (bit) | Throughput Normalized |
| 8 | 128 | 0.2724 | 128 | 0.2833 |
| 10 | 128 | 0.2812 | 160 | 0.2791 |
| 12 | 128 | 0.2823 | 192 | 0.3318 |
| 14 | 128 | 0.3467 | 224 | 0.4158 |
| 16 | 128 | 0.3601 | 256 | 0.5882 |

## 5.2 For CBC Mode of Cipher

Numerical results are obtained for the fixed and variable length encryption technique in Cipher Block Chaining mode of operation. In CBC mode of cipher plaintext is first XORed with initialization vector. Then encrypted data is fed to the next stage which is XORed with the next block of message to hide the message within the text or to hide the structure of the message. Simulation results for the cipher strength with fixed and variable block length encryption are shown in table 3. Graphical results for the comparative analysis of fixed and variable block length encryption are shown in figure 5. It is observed that the maximum optimal block length with channel adaptive block length encryption scheme at 256 bits is 133 bits and with fixed block length encryption it is 102 bits. These results show that link adaptive encryption scheme provides better security than fixed block length encryption scheme operating in CBC mode of cipher.

Table 3. Cipher Strength for Fixed and Variable Block Length Encryption in CBC Mode

| SNR (dB) | Fixed Block Length Encryption | | Variable Block Length Encryption | |
|---|---|---|---|---|
| | Block Length (bit) | Cipher Strength (bit) | Block Length (bit) | Cipher Strength (bit) |
| 8 | 128 | 22 | 128 | 85 |
| 10 | 128 | 73 | 160 | 116 |
| 12 | 128 | 86 | 192 | 107 |
| 14 | 128 | 54 | 224 | 113 |
| 16 | 128 | 102 | 256 | 133 |

Table 4. Throughput for fixed and variable block length of encryption in CBC mode

| SNR (dB) | Fixed Block Length Encryption | | Variable Block Length Encryption | |
|---|---|---|---|---|
| | Block Length (bit) | Throughput Normalized | Block Length (bit) | Throughput Normalized |
| 8 | 128 | 0.0072 | 128 | 0.0071 |
| 10 | 128 | 0.0079 | 160 | 0.0164 |
| 12 | 128 | 0.0940 | 192 | 0.0670 |
| 14 | 128 | 0.0939 | 224 | 0.0537 |
| 16 | 128 | 0.3989 | 256 | 0.4818 |

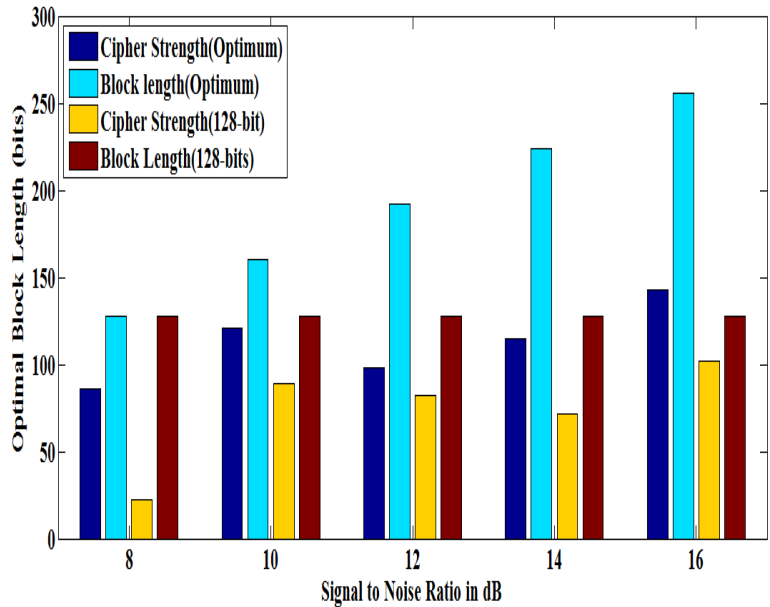

Figure 5. Cipher Strength for Fixed and Variable Block Length Encryption in CBC Mode

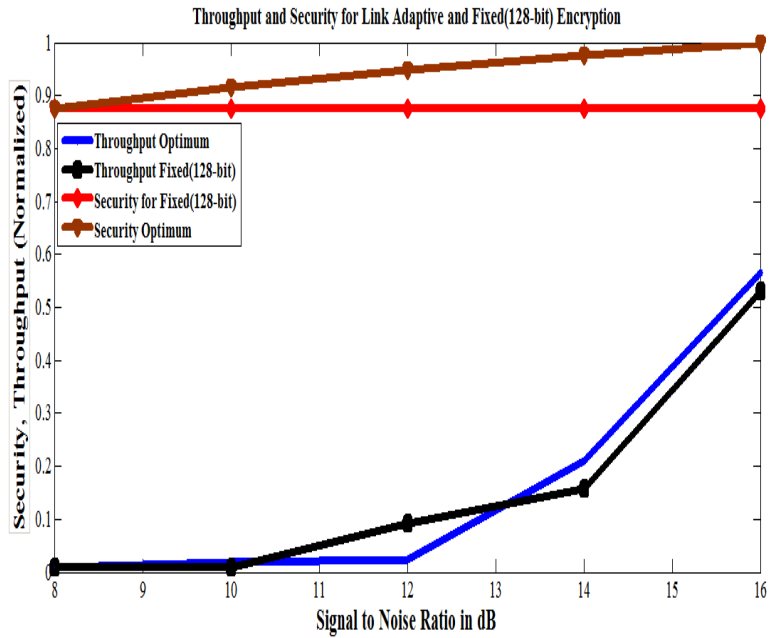

Figure 6. Normalized Throughput for Fixed and Variable Block Length Encryption with CBC Mode $S_{req}=.97$

Numerical results for the throughput are given in table 4. Normalized values of throughput have been calculated at different SNRs. There is an increase of 20-30% in throughput with link adaptive scheme as compared to that with fixed block length encryption scheme when operated in CBC mode of cipher. Graphical analysis for the comparison of throughput and security with fixed and variable block length in CBC mode of cipher is shown in figure 6. It is observed that link adaptive scheme is an optimized solution for security and throughput tradeoff as compared to fixed block length encryption scheme. Comparison of link adaptive scheme with a fixed block length encryption scheme operating in CBC mode of cipher has been performed which shows that Link adaptive scheme is proving better security and enhanced throughput as compared to fixed block length encryption scheme.

## 6. Conclusion

In this paper, we have addressed the problem of security-throughput trade-off in WLANs. Link Adaptive Encryption scheme based on the channel states was considered. Security is optimized in terms of brute force attack. Security of the cipher is measured by calculating the cipher strength or the optimal block length. Results were analyzed for the link adaptive encryption scheme and a fixed block length encryption scheme operating in two different modes of cipher.

Numerical results show that in the ECB mode, when SNR is 16dB, optimal block length at 256 bits is 198 bits for the variable block length encryption scheme and for fixed block length encryption the optimal block length at 128 bits is 221bits. It is observed that link adaptive encryption is providing very high level of security as compared to fixed block length encryption scheme. Throughput is also calculated for both variable and fixed block length encryption schemes. There is 70% increase in overall throughput with variable block length encryption scheme operating in ECB mode of cipher.

Numerical results were also obtained for fixed block length encryption and link adaptive scheme in CBC mode of cipher. It is observed that the optimal block length at 256 bits is 133 bits in case of variable block length encryption and at 128 bits is 102 bits for fixed block length encryption at 16dB SNR. It is observed that link adaptive scheme is more secure as compared to fixed block length encryption scheme. Throughput is increased by 20% in CBC mode of cipher for link adaptive scheme. Link adaptive encryption scheme yields enhanced security of a cipher without compromising the system performance and therefore results in optimized security throughput trade-off.

On comparing the link adaptive scheme operating in ECB and CBC modes, it is concluded that link adaptive encryption operating in CBC mode of cipher is the best suited encryption scheme for wireless network provided the channel states are known. Performance evaluation of link adaptive scheme for unknown channel conditions may be taken as the future direction of research work.

## 7. REFERENCES

[1] W. Stalling, "Cryptography and Network Security– Principle and Practice", Pearson Education, 2003.

[2] J.M. Reason, D.G. Messerschmitt, "The Impact of Confidentiality on Quality of Service in Heterogeneous Voice over IP Networks," Springer 2001, 2216, pp.175-192.

[3] S. Stein, "Fading Channel Issues in System Engineering," IEEE Journal on Selected Areas in Communications, 1987, vol 5, no. 2, pp. 68-89

[4] A. J. Goldsmith and Soon-Ghee Chua, "Variable-Rate Variable-Power MQAM for Fading Channels," IEEE Transactions on Information Theory, vol. 45, no.10, Oct. 1997, pp. 1218-1230.

[5] Rajput, S, "Wireless Security Protocols. In: Ilyas, M (ed.): Handbook of Wireless LANs," CRC Press, 2004.

[6] A. Stubblefield, J. Ioannidis, and A. Rubin, "Using the Fluhrer, Mantin, and Shamir Attack to Break WEP," AT&T Labs Technical Report, August 2001.

[7] Ezadin Barka, Mohammed Boulmalf, "On the Impact of Security on the Performance of WLAN," Journal of Communications, vol 2, no 4, pp 10-17, June 2007.

[8] Phongsak, Prasithsangaree and Prashant Krishnamurthy, " Analysis of Trade-off Between Security Strength and Energy Saving in Security Protocols for WLANs," Telecommunications Program, School of Information Science, University of Pittsburgh, 2004, pp. 5219-5233.

[9] Hanane Fathi, Kazukuni Kobara, Shyam S. Chakraborty, Hideki Imai and Ramjee Prasad, "Impact of Security on Latency in WLAN 802.11b," Proceeding of IEEE, Globecom 2005, pp. 1752-1756.

[10] Baghaei, Nilufar, "IEEE 802.11 Wireless LAN Security Performance Using Multiple Clients," Department of Computer Science and Software Engineering, University of Canterbury, Christchurch, New Zealand, 2004, pp 299-303

[11] Jian Liu, Jian Sun and Shoutao Lv., "A Novel Throughput Optimization Approach in Wireless Systems," IEEE 12th International Conference on Communication Technology (ICCT): 2010, pp. 1373-1377.

[12] M. A. Haleem, C.Nanjunda, and R. Chandramouli., "On Optimizing the Security-Throughput Trade-off in Wireless Networks with Adversaries," ACNS 2006, pp. 448-458.

[13] C.Nanjunda, M. A. Haleem and R. Chandramouli, "Robust Encryption for Secure Image Transmission over Wireless Channels," IEEE International Conference on Communications 2005, vol 2, pp. 1287-1291.

[14] Mohammed A. Haleem Chetan N. Mathur and R.Chandramouli, "Opportunistic Encryption: A Trade-off between Security and Throughput in Wireless Networks," IEEE Transactions on Dependable and Secure Computing, 2007, vol 4, no 4, pp. 313-324.

[15] W. C. Barker, "Recommendation for the Triple Data Encryption Algorithm (TDEA) Block Cipher," NIST Special Publication, pp. 800-67, NIST, May 2004.

[16] J. Daemen and V. Rijmen, "AES Proposal: Rijndael," http://csrc.nist.gov/CryptoToolkit/aes/rijndael/Rijndael.pdf, 2006.

[17] W. Trappe, L. Washington, "Introduction to Cryptography: With Coding Theory," Prentice Hall, 2002.

[18] M. Dworkin, "Recommendation for Block Cipher Modes of Operation: Methods and Techniques," NIST Special Publication, pp. 800-38A, NIST, 2001.

[19] McGrew, D., and Viega, J, "The Galois/Counter Mode of operation (GCM)," Submission to NIST, available from their web page, May 2005.

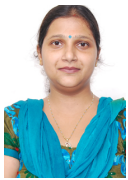
**Poonam Jindal** received B.E degree in Electronics and Communication Engineering from Punjab Technical University, Punjab in 2003, M.E degree in Electronics and Communication Engineering from Thapar University, Patiala in 2005 (India). She is working as Assistant Professor with Electronics and Communication Engineering Department, National Institute of Technology, Kurukshetra, India and currently pursuing her Doctoral Degree at National Institute of Technology, Kurukshetra, India**.** She has published seven research papers in International/National conferences. Her research interests include security algorithms for wireless networks and mobile communication. She is a member of IEEE.

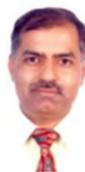
**Brahmjit Singh** received B.E. degree in Electronics Engineering from Malviya National Institute of Technology, Jaipur in 1988, M.E. degree in Electronics and Communication Engineering from Indian Institute of Technology, Roorkee in 1995 and Ph.D. degree from Guru Gobind Singh Indraprastha University, Delhi in 2005 (India). Currently, he is Professor, Electronics and Communication Engineering Department, National Institute of Technology, Kurukshetra, India. He teaches post-graduate and graduate level courses on wireless Communication and CDMA systems. His research interests include mobility management in mobile cellular networks, cognitive radio and wireless sensor networks. He has published more than 75 research papers in International/National journals and Conferences. Dr. Brahmjit Singh received the Best Research Paper Award from The Institution of Engineers (India) in 2006. He is a member of IEEE, life member of IETE and ISTE.